The acceleration of the expansion of the universe has been argued for by several research groups. If the universe is accelerating and if the universe or some part of the universe has a charge, then there may be electromagnetic radiation produced from the acceleration of the universe since accelerating charges produce electromagnetic radiation. This letter does a thought experiment to ask about the possible characteristics of the radiation from an accelerating universe. A value for the power, or rate of energy flow, of the universe's acceleration is calculated in this letter. The value of the power of the electromagnetic radiation emitted by the universe's acceleration is calculated to be $P = 5.99 \times 10^{-82}$ Joule/ sec or $5.99 \times 10^{-82}$ Watts. The letter then reveals what a characteristic of a charged, accelerating universe would be.
\\

Raymond Serway, in <u>Physics for Scientists and Engineers</u> 3rd edition, argues that "Accelerating charges produce EM radiation." "The fundamental mechanism responsible for this radiation is the acceleration of a charged particle. Whenever a charged particle undergoes acceleration, it must radiate energy." (Serway 1990, p. 970) If it is true that the expansion of the universe is accelerating (supported by research on 400 supernovae and two independent teams of researchers) and if it is true that the universe is charged or that some part of the accelerating universe has charge, then there should be a characteristic electromagnetic wave that radiates out from the acceleration of the universe. That is, the acceleration of a charged universe should produce a characteristic electromagnetic wave or radiation.

If we see the radiation at the predicted frequency or power, we have another piece of data for the acceleration of the universe. We also have a piece of data that some part of the universe is charged or that some part of the accelerating space-time is charged.

One could argue that the acceleration of the expansion of the universe is not the same as the kinetic acceleration of a charged particle, but still this letter does a thought experiment to ask about the characteristics of the radiation from an accelerating universe. If space-time's acceleration is the same as, or similar to, the acceleration of a particle, this letter will identify a characteristic of the electromagnetic radiation produced by the accelerating universe.

Now, what would the frequency and wavelength of that characteristic radiation be? A staff member at a planetarium pointed me towards the Larmor formula as a way to understand the radiation emitted by the acceleration of the universe. That staff member made it clear to me that this a "misuse" of the Larmor formula. The staff member argued that the Larmor formula refers to proper accelerations and not the kind that arise in an accelerating universe. The staff member also made clear to me that many, many observations of the universe have shown that we live in a universe described by the Friedman-Robertson-Walker (FRW) metric. That is, we live in a universe in which charged particles do not radiate in this FRW metric and in which there is local charge neutrality. The universe, the staff member argued, is known to be charge neutral to a very high degree.

Still, it is a useful exercise to determine what the power and frequency of this radiation might be if the universe were not described by the FRW metric and were not charge neutral. The exercise is useful because it tells us something about our universe. It tells us what we would see if the universe were charged or some part of the accelerating universe were charged. If we do not see radiation at this power or related frequency, it provides yet another piece of data that we live in a universe described by the FRW metric. If we do see electromagnetic radiation at this predicted power or frequency, then we have data that some part of the universe is charged. This letter to the editor then functions as a thought experiment. It asks, "What would be the power, or rate of energy flow, of the electromagnetic radiation produced by an accelerating universe?" It also asks for researchers to look for a predicted electromagnetic radiation from the acceleration of the universe.

So, what is a characteristic of the radiation from the accelerating universe?

We use--or "misuse"- the Larmor formula. The formula is--

The total power radiated is

P= e² a²/ 6 π (permittivity of free space) c³     (1)

This equation can be found in F. H. Read's <u>Electromagnetic Radiation</u>. (Read 1980, p. 54)

The values of these variables used for this letter follow.

Charge of electron as a possible value for the accelerating charge (e) = $1.602177 \times 10^{-17}$ C. Of course, the charge could be much stronger or much weaker.

Acceleration of space time (a) = With the assumption of a 100 km/sec increase in velocity for every million parsecs (mpc) Hubble relation, and assuming that the electrons or charged particles are moving at 100 km/sec, we get a value for the acceleration of the universe as $3.24 \times 10^{-16}$ km/sec²

Permittivity of free space= $8.854\ 187\ 817 \times 10^{-12}$ C²/Nm²

(c) or velocity of light = $2.998 \times 10^{8}$ m/ sec

Then, we get a value for the power, or rate of energy flow, of the electromagnetic radiation produced by the accelerating universe of

P = 5.99 x 10$^{-82}$ Joule/sec or 5.99 x 10$^{-82}$ Watts

Obviously, the power radiated is extremely, extremely small. As the staff member at a planetarium warned me before I determined this value, this is like looking for waves the size of an atom when there are waves four feet high at the beach. If there were very large charges accelerating in the expansion of the universe, the value of this rate of energy flow becomes larger. Still, the predicted power does have value, because it tells us about what the power of the radiation would be if the universe were indeed charged or some part of it were charged. The letter then reveals what a characteristic of a charged, accelerating universe would be.

Acknowledgements:

The idea of searching for a characteristic electromagnetic radiation from the acceleration of the universe came to me a few weeks before September 11, 2001 while reading the <u>Cartoon Guide to Physics</u>, a book Arno Penzias enjoyed. A staff member at Lowell Observatory helped me develop the value for the acceleration of space-time back in September 2001. A staff member at the Adler Planetarium helped me determine that the Larmor formula is what I needed to use to turn the acceleration of the universe into a value for the radiation emitted.

I hope that this analysis or thought experiment might be useful to some scientist, researcher, or someone interested in astronomy. I would also like to acknowledge Albert Einstein for making me feel ok about doing a thought experiment when a few people called it a silly idea.

## Reference List

Read, F. 1980, Electromagnetic Radiation (Chichester, John Wiley & Sons)

Serway, R. 1990, Physics for Scientists and Engineers (3rd Edition, Philadelphia: Saunders Golden Sunburst Series)